\documentclass[%
 amsmath,amssymb,amsfonts,aps,
]{revtex4-1}
\pdfoutput=1
\usepackage{hyperref}
\usepackage{graphicx}
\usepackage{dcolumn}
\usepackage{bm}
\usepackage{amssymb}

\usepackage[utf8]{inputenc}
\usepackage[T1]{fontenc}
\usepackage{mathptmx}

\begin{document}

	\title{Constraining Model Uncertainty in Plasma Equation-of-State Models with a Physics-Constrained Gaussian Process}
	\thanks{LLNL-JRNL-835975}

	\author{Jim A. Gaffney}
	\email{gaffney3@llnl.gov}
	\author{Lin Yang}
	\author{Suzanne Ali}
	\affiliation{Lawrence Livermore National Laboratory, 7000 East Ave, Livermore, CA 94550, USA}

	\date{\today}





	\begin{abstract}
Equation-of-state (EOS) models underpin numerical simulations at the core of research in high energy density physics, inertial confinement fusion, laboratory astrophysics, and elsewhere. In these applications EOS models are needed that span ranges of thermodynamic variables that far exceed the ranges where data are available, making uncertainty quantification (UQ) of EOS models a significant concern. Model uncertainty, arising from the choice of functional form assumed for the EOS, is a major challenge to UQ studies for EOS that is usually neglected in favor of parameteric and data uncertainties which are easier to capture without violating the physical constraints on EOSs. In this work we introduce a new statistical EOS construction that naturally captures model uncertainty while automatically obeying the thermodynamic consistency constraint. We apply the model to existing data for $B_4C$\ to place an upper bound on the uncertainty in the EOS and Hugoniot, and show that the neglect of thermodynamic constraints overestimates the uncertainty by factors of several when data are available and underestimates when extrapolating to regions where they are not. We discuss extensions to this approach, and the role of GP-based models in accelerating simulation and experimental studies, defining portable uncertainty-aware EOS tables, and enabling uncertainty-aware downstream tasks.
	\end{abstract}


	\maketitle

%
\section{Introduction}

Understanding high energy density physics, inertial confinement fusion and laboratory astrophysics experiments relies heavily on equation-of-state (EOS) models due to their importance in closing the equations of hydrodynamics\cite{Gaffney_2018}. In these applications different portions of the system pass through a range of thermodynamic states that greatly exceeds that accessible by focussed EOS measurements. As a result EOS models routinely extrapolate away from the data used to benchmark them, making model validation and uncertainty quantification a significant concern.\par
Typical EOS models are built by collecting a patchwork of experimental data and first-principles simulations of different types and fitting a parametric function that can be used to interpolate and extrapolate over a wide range of input states\cite{Militzer_2021, Zhang_2020}. This parametric approach is attractive since it allows important physical constraints and limiting behaviors to be trivially enforced by picking a physically-motivated functional form\cite{Wu_2019}. Uncertainty in the resulting EOS can be measured by statistical sampling of the fitted parameters, and propagating to the quantities of interest. This approach captures the data component of the total EOS uncertainty but misses another: the uncertainty in the choice of functional form used, which we call \emph{model uncertainty}. In this paper we report on work to constrain model uncertainty in equation-of-state models taking into account the physics-driven constraints on the model form.\par
Constraining model uncertainty involves exploring the potentially infinite space of viable functional forms. This can be approximated by considering an ensemble of parametric functions and fitting each individually, or by fitting a highly flexible non-parametric model that implicitly represents an entire class of functions. The first approach has the advantage that every member of the ensemble can be a physically viable form, but risks underestimating the uncertainty since it is unlikely to consider all possible functions; the second approach allows for a much more complete quantification of the uncertainty but efforts must be made to ensure the space of functions explored is physically reasonable.\par
In this work we develop a non-parametric fitting procedure suitable for EOS applications. Our approach is based on a variation of the well-known Gaussian process (GP) regression \cite{Mackay_1998}, where we introduce modifications to ensure that all functions generated by the GP obey the laws of thermodynamics and that the known ideal behavior at high temperature is recovered. We demonstrate our model by applying it to published data for Boron Carbide ($B_4C$), and use it to explore the uncertainty in pressure $P$ and internal energy $E$ when interpolating first-principles simulation data.\par
Uncertainty quantification of materials properties is being actively researched by several groups. Recently, Ali \emph{et al.}\cite{Ali_2020} demonstrated a full quantification of parametric uncertainty in the EOS of copper by Monte-Carlo sampling data uncertainties and propegating through the EOS fitting process. Kamga \emph{et al.}\cite{Kamga_2014} have quantified the uncertainty in a single model by exploring the discrepancy with experiment. Root \emph{et al.} \cite{Root_2010}, Walters \emph{et al.}\cite{Walters_2018}, and Brown \emph{et al.}\cite{Brown_2018} have used Gaussian process based model calibration \cite{Kennedy_2001, Higdon_2008} to update physics models to better agree with experiment. While important steps, none of these approaches fully acount for the range of models that are consistent with experimental data, or constrain the calibrated model to be consistent with fundamental physics. This is the main contribution of our work.\par

%
\section{Model Uncertainty in Equations-of-State}

A general EOS model takes a pair of natural variables as input and predicts all other thermodynamic variables of the system as outputs, taking into account phase changes, molecular and atomic physics, coupling with a radiation field, etc. In a hydrodynamics simulation, the input variables are those evolved on the computational mesh, typically the temperature $T$ and mass density $\rho$ or volume $V$; the outputs are pressure, internal energy, entropy, specific heat, and any other material property of interest. In this way the EOS summarizes the microscopic response of the material with the aim of allowing the complex sub-grid physics to be modeled separately from the hydrocode itself.\par
The usual process of building an EOS model is to use as much data as possible to learn a mapping from the input space $\mathbf{x}\in\mathbb{R}_{>0}\times\mathbb{R}_{>0}$ into the outputs $\mathbf{y}\in\mathbb{R}^N$. The data are usually a combination of experimental results from a variety of facilities and platforms, sometimes combined with results from multiple first-principles simulation studies with varying fidelities and regions of applicability. The resulting model may be written
\begin{equation}
EOS: \mathbf{x},\mathbf{\alpha}\mapsto\mathbf{y}
~{\rm ,}
\label{eq:eos_mapping_1}
\end{equation}
where $\mathbf{\alpha}$ is a vector of parameters whose values are determined by fitting to the available data over the entire range of the EOS.\par
Quantifying our confidence in the EOS in \eqref{eq:eos_mapping_1} involves admitting that, for a given input points $x$, there may be a range of outputs which are consistent with the available data. The variations come from two sources: there may be multiple possible mappings (model uncertainty); and for a fixed mapping there may be multiple sets of parameters $\alpha$ (parametric uncertainty). These two contributions may contribute equally to the total uncertainty in the predicted quantities, however capturing model uncertainty is more challenging since it involves enumerating all of the potential mappings (functional forms) that could be valid EOSs. As a result of this difficulty, most previous work has considered only parametric uncertainty and can be expected to underestimate the total uncertainty in the EOS model.\par
In the remainder of this sections we will present a method for automatically exploring a continous space of EOS models to generate predictions that capture all sources of uncertainty. We being by defining the function space in terms of basic physical constraints, and then present a Gaussian process method for sampling general EOSs conditional on data.\par
%
%
\subsection{Thermodynamic Constraints on EOS Functional Forms}
The outputs of a physically realistic EOS model are strongly constrained by fundamental thermodynamics. On such constraint, known as \emph{thermodynamic consistency}, arises from the fact that \emph{all} output quantities - the elements of $\mathbf{y}$ - may be derived directly from the thermodynamic potential, a single scalar function of the inputs $\mathbf{x}$. These relationships give rise to well-known differential relationships which can be used to define the space of `reasonable' EOS models. In this section, we will recap one common form of the consistency constraint.\par
In hydrocode applications it's common to use $\mathbf{x}=(T,V)^T$, where $T$ is the temperature and $V$ is the volume. In that case, the relevant thermodynamic potential is the Helmholtz free energy,
\begin{equation}
F = U - TS
~{\rm ,}
\label{eq:helmholtz_free_energy}
\end{equation}
where $U$ is the internal energy and $S$ is the entropy. Combining with the fundamental thermodynamic relation,
\begin{equation}
dU = -SdT - PdV
~{\rm ,}
\label{eq:fundamental_thermodynamic_relation}
\end{equation}
where $P$ is the pressure, allows the EOS quantities $(P,U,S)$ to be defined in terms of $F$;
\begin{gather}
\begin{align}
P &= -\left.\frac{\partial F}{\partial V}\right\vert_T
~{\rm ,} \\
U &= F - T\left.\frac{\partial F}{\partial T}\right\vert_V
~{\rm ,} \\
S &= -\left.\frac{\partial F}{\partial T}\right\vert_V
~{\rm .}
\end{align}
\label{eq:ops}
\end{gather}
Finally, differentiating the first two equations by $T$ and $V$ respectively gives the thermodynamic constistency constraint between EOS ouputs,
\begin{equation}
P=T\frac{\partial P}{\partial T} + \frac{\partial U}{\partial V}
~{\rm .}
\label{eq:td_consistency}
\end{equation}
\par
Obeying the consistency relation \eqref{eq:td_consistency} is often considered a fundamental requirement for a useful EOS model, and has been used to test the numerical convergence of first-principles simualtions. The space of functions that obey the relation is, therefore, the maximal set of functions that may be used for EOS work, and may be used to provide an upper limit on model uncertainty. In the next section, we present a method for sampling from the space of functions that identically satisfy equation \eqref{eq:td_consistency}.\par
\subsection{A Thermodynamically Constrained Gaussian Process}\label{sec:thermo_gp}
Gaussian processes (GPs) are a popular method for uncertainty quantification since since they are lightweight and have a clear statistical interpretation, resulting in a natural description of prediction uncertainties. They are also highly flexible meaning that they can represent a wide range of functional forms in a single closed form. This interpretability and flexibility makes GPs a popular choice when quantifying model uncertainty.\par
Another property of GPs of particular relevance here is that a linear transformation of a GP is another GP. This is useful result when working with constrained functions, since a weakly constrained GP of standard form can be transformed into a strongly constrained one by applying suitable operators. This approach has recently been used to build GPs that automatically solve differential equations\cite{Lange_2018} and to enforce a curl-free vector output suitable for fitting magnetic fields\cite{Jidling_2017}, and is the approach we take in this work.\par
%
%
A Gaussian process models the joint distribution of function outputs as a multivariate normal distribution. The correlation between output values is determined by their relative input values - typically, points closer in input space will be more strongly correlated in output space. The details of this correlation structure is controlled by the GP kernel, a scalar function which determines the correlation between two input points $\mathbf{x}$ and $\mathbf{x}^\prime$. At the core of this work is a GP kernel that enforces correlations between EOS quantities and their derivatives in such a way as to guarentee thermodynamic consistency.\par
%
%
We start by introducing a general GP for the Helmholtz free energy,
\begin{equation}
F\sim GP(F_R(\mathbf{x}), k_F(\mathbf{x},\mathbf{x}^\prime))
~{\rm ,}
\label{eq:gp_F}
\end{equation}
where we have introduced a reference free energy $F_R$ as the prior mean, and $k_F$ is the free energy kernel. For further details on GPs and their implementation, we refer the reader to any textbook or tutorial on the subject\cite{Schulz_2018, uqpy_gps, stan_gps}. For our purposes here, it suffices to note that this model for $F$ is unconstrained except for a weak prior on the smoothness of the functional form introduced by the kernel, and the reference free energy that determines the long-range extrapolation behavior of the free energy.\par
The next stage in our analysis is to transform from the unconstrained free energy to the constrained space of EOS outputs. Using equations \eqref{eq:ops} to define a linear operator $\mathcal{\hat{Y}}_{\mathbf{x}}$ so that $(P,U,S)^T = \mathcal{\hat{Y}}_{\mathbf{x}}F(\mathbf{x})$, we can use a standard identity for the action of a linear operator on a multivariate normal to define the multi-output EOS GP (see appendix \ref{sec:appendix_A} for some mathematical details),
\begin{gather}
\begin{pmatrix} P \\ U \\ S \end{pmatrix} \sim GP(\mathcal{\hat{Y}}_{\mathbf{x}}F_R, \mathcal{\hat{Y}}_{\mathbf{x}^\prime}k_F\mathcal{\hat{Y}}_{\mathbf{x}}^T)
~{\rm ,}
\label{eq:eos_gp}
\end{gather}
wheere we have supressed the dependance on input $\mathbf{x}$ except where necessary to clarify the argument of the kernel function on which $\mathcal{\hat{Y}}$ acts. The transformed GP \eqref{eq:eos_gp} is constrained to sample functions with the correct thermodynamic consistency relations. Given a choice of reference and kernel functions for the free energy, the mean and kernel for the EOS model can be found through straightforward manipulations, easily handled by a symbolic algebra code. Finally, since any thermodynamic quantity can be written as a linear transformation of the free energy, our model is easily generalized by adding the relevant rows to the operator $\mathcal{\hat{Y}}$. \par
Equation \eqref{eq:eos_gp} provides a general approach for assessing uncertainty and confidence in EOS models and tables. The trained GP efficiently represents a continous space of EOS models that are consistent with the data; generating multiple tables for use in hydrocodes can be done simply by sampling from a single multivariate normal distribution. Finally, since GPs are a popular tool in all areas of uncertainty quantification, this picture opens the door for a multitude of new developments including uncertainty propegation, acceleration of simulation studies, and principled design and analysis of new experiments.\par

%
\section{Application to high-temperature Boron Carbide}
The development of new theoretical EOS models typically proceeds by running a large set of first-principles simulations and using them to train a wide ranging model. One example is the recent work for $B_4C$ by Zhang \emph{et al.}\cite{Zhang_2020}, which combined simulations using path integral Monte Carlo (PIMC), quantum molecular dynamics (QMD), Green’s function Korringa-Kohn-Rostoker (KKR) theory, activity expansion (ACTEX) and other methods to build a wide ranging interpolation which was cross-validated by experiments at the National Ignition Facility. For the purposes of this paper, we will use the results of the PIMC and QMD simulations which form a set of 231 samples of the mapping $(T, \rho)\rightarrow(P, U)$, with statistical uncertainties on $(T, P, U)$. Note that for consistency with Zhang \emph{et al.}we will plot the mass density $\rho$, which is related to the volume used in section \ref{sec:thermo_gp} by $\rho({\rm g/cc}) = 619.8 / V(a^3_0/{\rm atom})$, and we do not include the entropy output in equation \eqref{eq:eos_gp} since those values were not reported. We use these data to find optimal hyperparameters for our EOS GP regression model as described previously, and to make predictions of the uncertainties in the EOS.\par
%
%
%
\begin{figure*}
\centering
\includegraphics[scale=1]{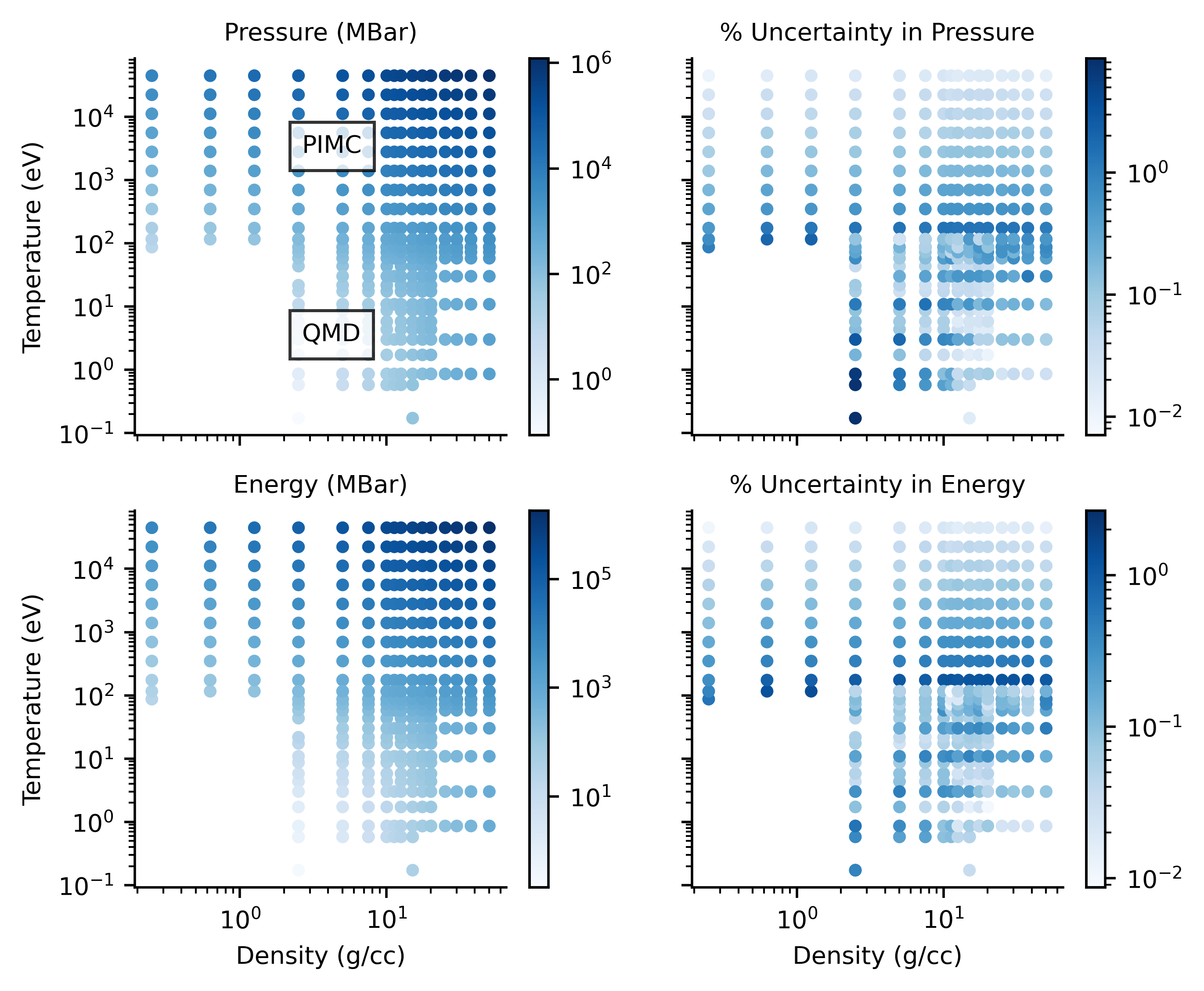}
\caption{First-principals data used to train our thermodynamically consistent Gaussian Process, consisting of previously-published path integral Monte Carlo (PIMC) and quantum molecular dynamics (QMD) simulations for boron carbide\cite{Zhang_2020}. We use the published pressure $P$ and internal energy $U$, as a function of temperature $T$ and mass density $\rho$, along with statistical uncertainties on $T$, $P$ and $U$. The two simulation methods overlap in a small region around $T=100 {\rm eV}$, where both are pushed to the limits of their applicability.}
\label{fig:simulation_data}
\end{figure*}
We plot the pressure $P$ and internal energy $U$, along with their quoted uncertainties, in figure \ref{fig:simulation_data}. While the EOS quantities themselves are in good agreement between the PIMC and DFT methods, the uncertainties are quite different and show a discontinuous change where the two methods overlap at $T\sim100 \rm{eV}$. In the PIMC simulations statistical uncertainties grow to almost 10\% at lower temperatures while the QMD uncertainties are much lower. One goal of our GP regression technique is to combine the results from both simultions and provide a measure of the true uncertainty in this region.\par

The data in figure \ref{fig:simulation_data} also identify a practical challenge to fitting GPs, and machine learning models in general, to EOS data: the inputs and outputs must span many orders of magnitude. For the inputs $T$ and $V$ we can take the standard approach and log-scale, however our method of enforcing constraints requires that any transformation of the outputs $P$ and $U$ is linear, making log-scaling inappropriate for those quantities. We instead rely on the reference model to capture the large-scale changes in output and use the GP to model deviations around the reference, known as the \emph{excess} pressure $\Delta P$ and internal energy $\Delta U$. For this dataset, the excess EOS quantities vary smoothly between 0.5 and 3, a range which is suitable for reliable training of the GP model.\par
\subsection{Specifying and Training the EOS GP}
Our model is specified by the reference free energy $F_R$ and free energy kernel function $k_F$. The choice of reference function does not effect the behavior of the GP close to training data, but does determine the functional form on extrapolation since the posterior mean reverts to the reference in this limit. Here, we will use the ideal gas free energy. The choice of kernel function is also relatively free and is intuitively related to our prior belief on the form of $F$. In this work, we use a Gaussian or radial basis function (RBF) kernel,
\begin{gather}
k_F(lnT, ln\rho; lnT^\prime, ln\rho^\prime) = \sigma e^{-\frac{(lnT-lnT^\prime)^2}{2\lambda_{lnT}^2}} e^{-\frac{(ln\rho-ln\rho^\prime)^2}{2\lambda_{ln\rho}^2}}
~{\rm ,}
\label{eq:rbf_kernel}
\end{gather}
which introduces an unknown correlation length for each input direction, $\lambda_{lnT}$ and $\lambda_{ln\rho}$, and an overall correlation strength $\sigma>0$. These parameters are learned from the EOS data based on the probability of the data defined by equation \eqref{eq:eos_gp}. For this work, the optimized correlation lengths are of particular interest since they describe the lengthscale over which prediction uncertainty increases as the GP is extrapolated away from available data.\par
Given the choice of free energy kernel function \eqref{eq:rbf_kernel} we use the computer algebra package \texttt{sympy} to determine the thermodynamically consistency kernel function. Mathematical details and a description of the resulting kernel are given in appendix \ref{sec:appendix_A}. With the kernel function in hand we `train' the EOS GP by optimizing the parameters $\sigma$, $\lambda_{lnT}$ and $\lambda_{lnV}$ by maximizing the log-probability of the data. It is well known that the training process can run into numerical issues when data are close together and have small uncertainties, as is the case here; we solve this problem by randomly perturbing the training inputs according to the statistical uncertainties in the simulated temperatures, and using a stochastic gradient descent algorithm that is robust to these variations. We have found that this is a reliable regularization that does not require \emph{ad-hoc} modifications to the GP model, and has the added bonus of providing a relatively simply way of accounting for uncertainties in the simulated temperature. At the end of this process we find optimal parameters $\lambda_{lnT}=4.78$, $\lambda_{ln\rho}=4.03$, for which the $R^2$ correlation coefficient between simulation data and GP prediction is better that $99.5\%$.\par
\subsection{Uncertainties in the $B_4C$ EOS}
\begin{figure*}
\centering
\includegraphics[scale=1]{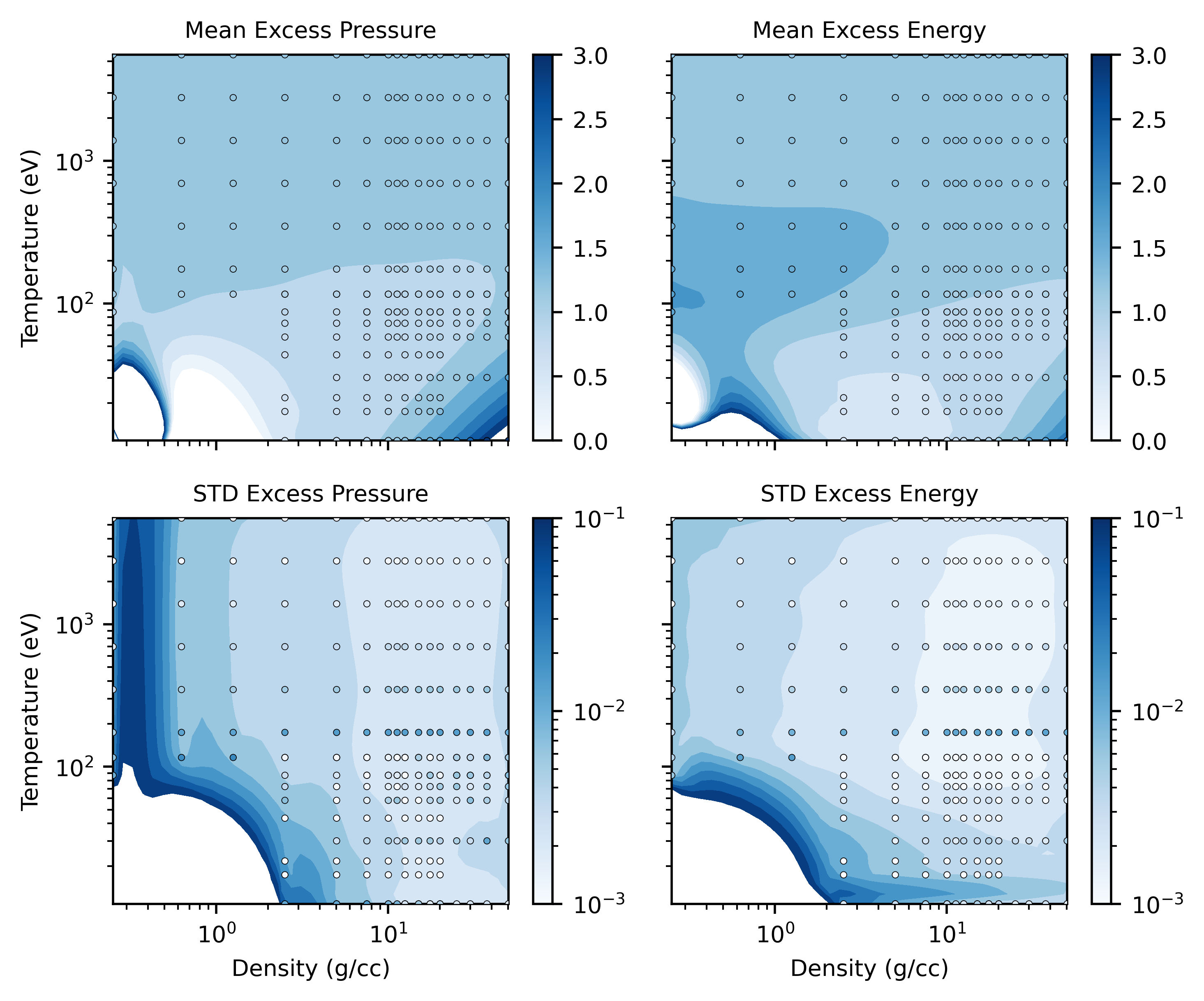}
\caption{Predictions from our constrained Gaussian process EOS model for $B_4C$. Points show the simulation values (top row) and uncertainties (bottom row) used to train the GP, and contours show the GP interpolation, on a common colorscale. The thermodynamically consistent GP faithfully reproduces the training data ($R^2 = 0.995$) with a significantly reduced uncertainty.}
\label{fig:predicted_eos}
\end{figure*}

The optimized EOS GP, for the first time, provides us with a wide ranging EOS model that captures both data and model uncertainty. We visualize the uncertain model by plotting the mean and standard deviation of the predicted pressure and energy as a function of input density $\rho$ and temperature $T$ in figure \ref{fig:predicted_eos}.\par
The prediction means in the top row of \ref{fig:predicted_eos} compare the mean of the predicted excess pressure and energy (contours) with the simulations (points) on a common color scale. At the locations of the simulation points the GP mean is in excellent agreement with the simulations, with an average error of $1\%$, and the GP prediction uncertainty is typically a factor of $10$ lower than the uncertainty in the simulation themselves. Away from the training data the EOS GP picks up important features from the simulations and smoothly interpolates over the whole space. In principal the EOS GP model can predict the pressure and energy at \emph{any} value of $lnT$ and $ln\rho$, however prediction uncertainties rapidly increase away from the available data as shown in plots of the prediction uncertainty in the bottom row of figure \ref{fig:predicted_eos}. The rate of increase is a combination of the inferred smoothness of the free energy, described by the optimized values of $\lambda_{lnT}$ and $\lambda_{ln\rho}$, and the effect of the constraints. We find that while the inferred length scales suggest that the free energy can be extrapolated for a decade or more in $T$ and $\rho$, uncertainty in the EOS qantities grows much faster as a result of the thermodynamic constraints.\par
\begin{figure}
\centering
\includegraphics[scale=1]{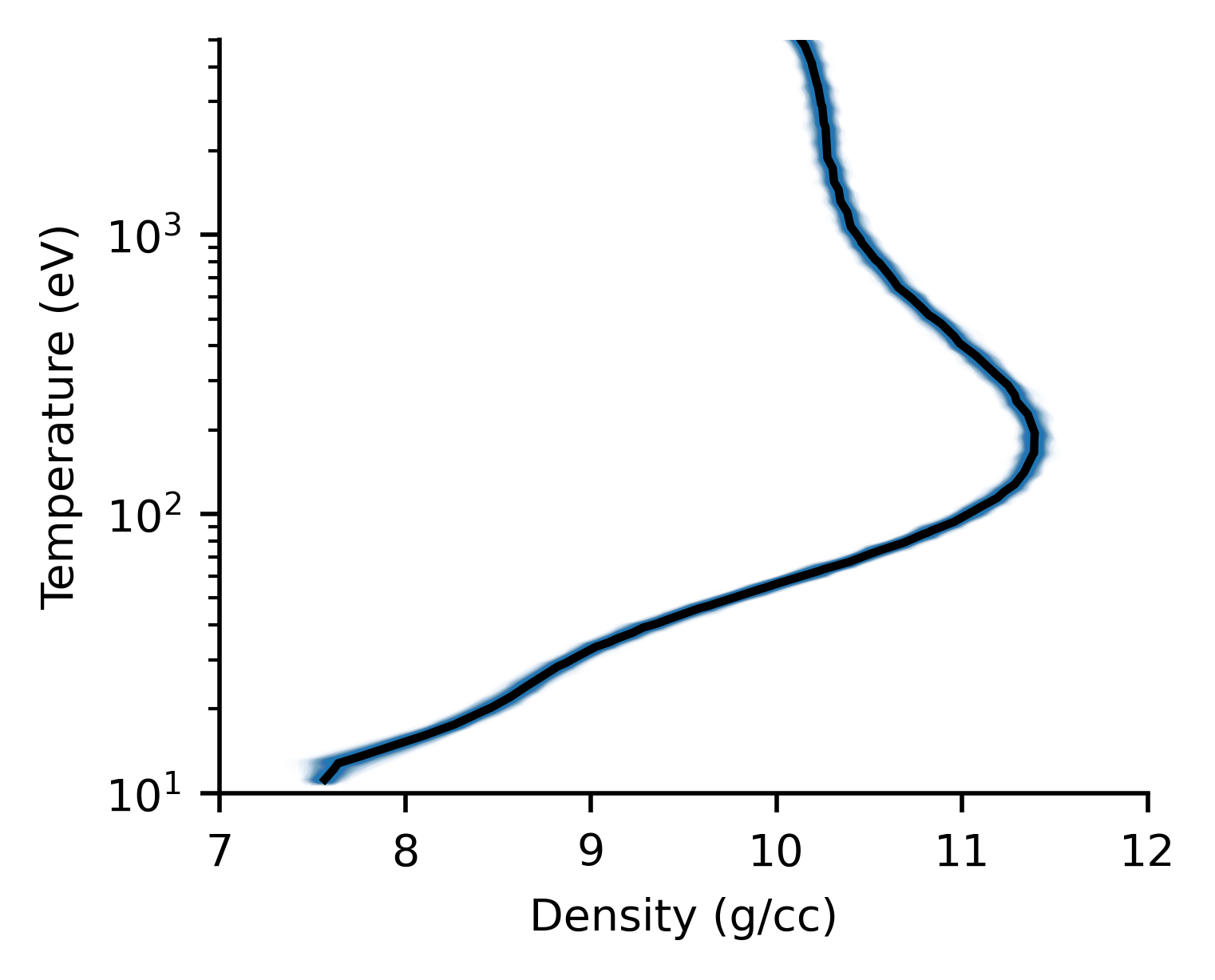}
\caption{Predicted Hugoniot, with uncertainty, for $B_4C$. We use our GP model to sample from the distribution of equation-of-state models that are consistent with our first-principles simulations. For each EOS sample, we calculated the Hugoniot curve and plot the standard devition in density at constant temperature.}
\label{fig:predicted_hugoniot}
\end{figure}

Our GP model provides a principled and efficient way of sampling from the space of EOS models for downstream tasks, for example in hydrodynamics simulations or the analysis of experimental data. One important example is the calculation of the \emph{Hugoniot}, the set of thermodynamic states that can be accessed through shock compression of a material, along which the majority of experimental data are constrained to lie. We show the distribution of shock Hugoniot curves for $B_4C$ in figure \ref{fig:predicted_hugoniot}, generated by sampling from our GP model and postprocessing each sample to generate a set of Hugoniot curves. The uncertainty in the Hugoniot, measured here by the standard deviation in predicted density at given temperature, is lower than the uncertainty in the underlying EOS quantities as a consequence of the dense sampling of simulations in the Hugoniot region and the physically correlated uncertainties in EOS quantities that our model provides.\par

%

\subsection{Importance of the thermodynamic consistency constraint}

Our EOS GP model ensures thermodynamic consistency by limiting the space of functions that it can sample. A result of this is that our GP model can be expected to have lower prediction uncertainties than a standard unconstrained on. In effect, the laws of thermodynamics have been used as an additional source of information resulting in lower uncertainty.\par
\begin{figure*}
\centering
\includegraphics[scale=1]{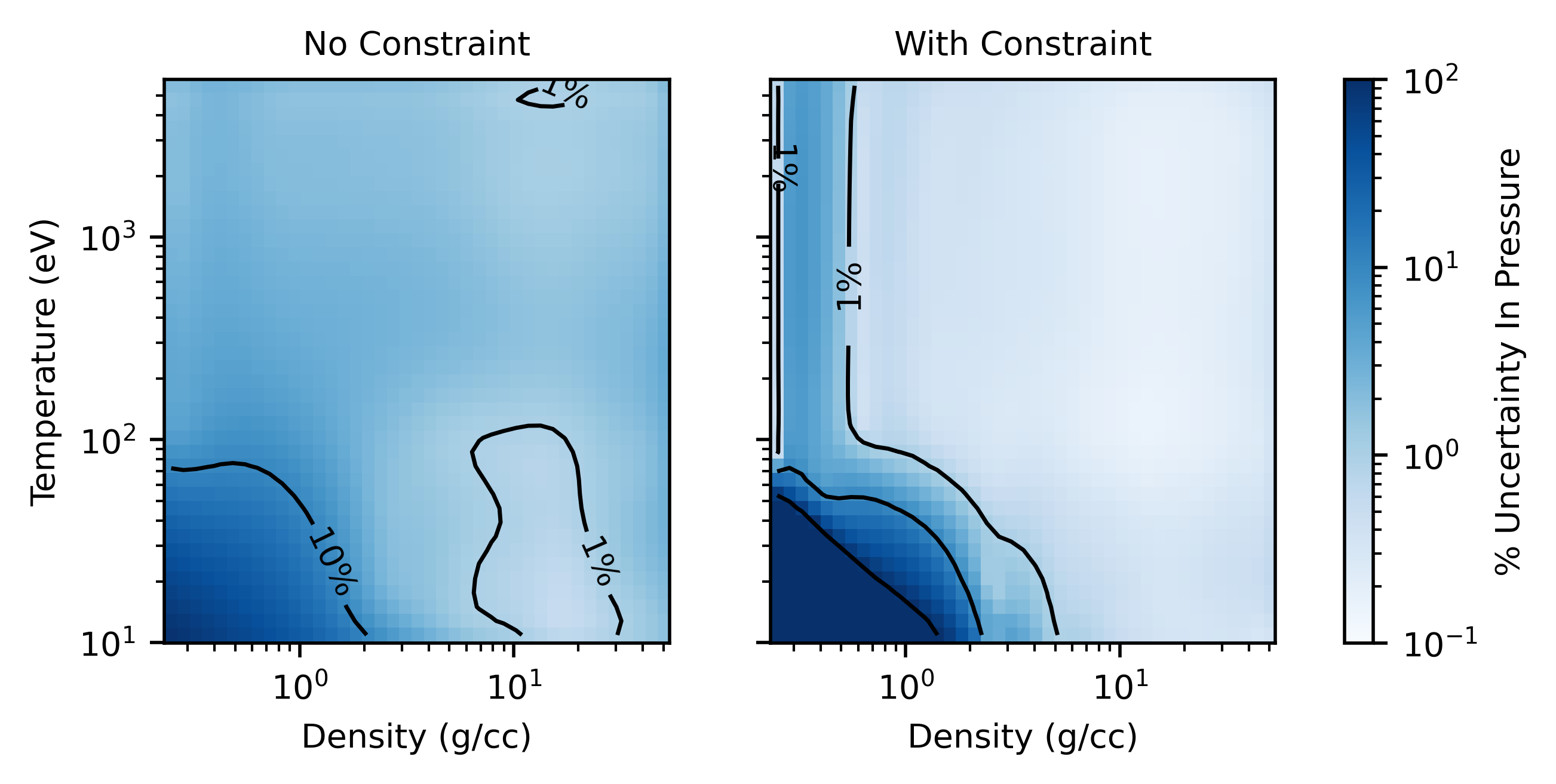}
\caption{Including the thermodynamic consistency constraint on model form significantly reduces the uncertainty in the resulting EOS. We compare the uncertainty in the predicted pressure for $B_4C$ from a Gaussian process models that neglect (left) and include (right) the constraint. Inclusion of the constraint reduces the uncertainty by a factor of 10 over most of the space where training simulations are available.}
\label{fig:constraint_comparison}
\end{figure*}
We compare uncertainties with and without the thermodynamic constraint by plotting the fractional uncertainty in predicted pressure from two models in figure \ref{fig:constraint_comparison}. In the figure, the left panel shows the uncertainty in pressure for a GP with a standard RBF kernel which samples from a space of smooth functions but does not enforce any other constraints; the right panel shows the results from our EOS GP model. The effect of the constraint is clear; for the standard GP the prediction uncertainty falls below 1\% only where simulations are densely sampled and the temperature is low enough that statistical noise on the DFT simulations is low, while for the EOS GP the uncertainty is less than 1\% over almost the entire range of the simulated dataset. In the extrapolation region at the bottom left of each plot the effect of the constraint is also clear, in increasing uncertainty as discussed in the previous section. Clearly, an approach to model uncertainty that neglects the constraints would dramatically overestimate uncertainty where data are available, and underestimate when extrapolating into regions where no data can be collected. \par
%
%
\section{Summary \& Further Work}
In this work we have developed a thermodynamically constrained Gaussian process for use in equation-of-state (EOS) modeling. Our EOS GP is constrained to explore the space of thermodynamically consistent functions by explicitly relating all thermodynamic quantities to an underlying free energy model. When trained on simulation or experimental data, the EOS GP naturally estimates model uncertainty in the predicted thermodynamic quantities which, until now, have been very difficult to capture. This is the main contribution of this work so far. \par
We have demonstrated the importance of model uncertainty, and the effect of the thermodynamic constraints, by applying our model to existing first-principals simulations for $B_4C$. We show that our EOS GP provides a high quality representaion of the EOS with uncertainties of less than $1\%$ over a wide range of temperatures and densities, and that prediction uncertainties are generally much lower than the statistical uncertainties in the simulations themselves. Comparison with an unconstrained GP model shows that neglecting the constraint results in an \emph{overestimate} of the uncertainty by factors of several in regions where data are available, and an \emph{underestimate} when extrapolating to new parts of the parameter space. As a result, an attempt to measure model uncertainty without accounting for the thermodynamic constraints is likely to be unreliable.\par
The thermodynamic consistency constraint considered here is not the only constraint on EOS models, and as a result the uncertainties we have presented are an upper bound. Perhaps the most important additional constraint is thermodynamic stability resulting in positivity of the specific heat and bulk modulus. Since these are inequality constraints enforcing them requires a different approach to the one adopted here\cite{Swiler_2020}, and will be the subject of further work.\par
Known limiting behaviour is also an essential component of a meaningful EOS model. Our EOS GP controls long-range behaviour through the introduction of a reference free energy as the prior mean which ensures that the mean prediction reverts to the reference far from any available data. However, in the current approach uncertainty will (incorrectly) continue to grow while the model reverts to the known behaviour. Controlling the increase in uncertainty, as well as implementing multiple material phases, in order to provide a sufficiently wide-ranging EOS to be of use in hydrodynamics simulations is ongoing.\par
Despite the further work needed, the use of Gaussian processes has already provided important insights, and promises to significantly change the way the EOS tables are generated and stored. Prediction uncertainties can be used to adaptively choose points for new simulations or experiments to focus on uncertain regions, potentially reducing the number of data needed and accelerating the development of new EOSs. GPs also provide a natural and memory-efficient representation of the full distribution of EOS models in a single object, and do not require stuctured grids in $T$ and $\rho$, which suggests a useful extension to the current EOS table formats to include uncertainty. As a result we believe that GPs could form the core of a new uncertainty-aware EOS modeling framework.\par
%
%
\begin{acknowledgments}
Prepared by LLNL under Contract DE-AC52-07NA27344.
\end{acknowledgments}


\bibliography{ref}

%
\appendix
\section{Mathematical Details}
\label{sec:appendix_A}
Our goal is to derive a Gaussian process kernel that results in a statistical model that samples the space of thermodynamic functions. In this section we will provide some mathematical details and explicit details on the approach taken to derive the kernel used in this work. Much of the details here follow the work of Jidling \emph{et al.}\cite{Jidling_2017} with some specializations for the the application to EOS modeling. Finally we will make some general observations about the structure of the EOS GP kernel. \par
\subsection{General Approach}
We start by defining a free energy
\begin{equation}
F:\mathbb{R}^2\mapsto\mathbb{R};~\mathbf{x}\mapsto F(\mathbf{x})
\end{equation}
and an equation of state
\begin{equation}
EOS:\mathbb{R}^2\mapsto\mathbb{R}^N;~\mathbf{x}\mapsto \mathbf{y}(\mathbf{x})
~{\rm ,}
\end{equation}
both of which act on a pair of natural thermodynamic variables $\mathbf{x}$. As discussed in the main text, the choice of natural variables defines the form of $F$ and the laws of thermodynamics give operators relating free energy to EOS outputs, which we write
\begin{equation}
\mathcal{\hat{Y}}_\mathbf{x}:\mathbb{R}\mapsto\mathbb{R}^N;~F(\mathbf{x})\mapsto\mathcal{\hat{Y}}_\mathbf{x}F(\mathbf{x}) =
\begin{pmatrix}
  \mathcal{\hat{Y}}_\mathbf{x}^{(1)}F(x) \\
  \vdots\\
  \mathcal{\hat{Y}}_\mathbf{x}^{(N)}F(x)
\end{pmatrix}
~{\rm .}
\end{equation}
\par
A key observation for this work is that each of the $\mathcal{\hat{Y}}_\mathbf{x}^{(i)}$ is a linear operator, in which case introducing a Gaussian process for the free energy $F(\mathbf{x})\sim GP(F_R(\mathbf{x}),\,K_F(\mathbf{x}, \mathbf{x}^\prime))$ results in another GP for the EOS quantities with a transformed mean and covariance function,
\begin{equation}
  y(\mathbf{x})\sim GP(\mathcal{\hat{Y}}_\mathbf{x}F_R(\mathbf{x}),\,\mathcal{\hat{Y}}_{\mathbf{x}^\prime}K_F(\mathbf{x}, \mathbf{x}^\prime)\mathcal{\hat{Y}}_\mathbf{x}^T)
  ~{\rm .}
  \label{eq:eos_gp_app}
\end{equation}
As a result of the transformation to EOS outputs, the free energy kernel has become an $N\times N$ matrix of kernel functions with elements
\begin{equation}
  (\mathcal{\hat{Y}}_{\mathbf{x}^\prime}K_F(\mathbf{x}, \mathbf{x}^\prime)\mathcal{\hat{Y}}_\mathbf{x}^T)_{ij} = \mathcal{\hat{Y}}_{\mathbf{x}^\prime}^{(i)}\, \mathcal{\hat{Y}}_{\mathbf{x}}^{(j)}\,K_F(\mathbf{x}, \mathbf{x}^\prime)
  ~{\rm .}
  \label{eq:eos_kernel_elements}
\end{equation}
The off-diagonal kernel functions, $i\ne j$, play a critical role in our model by enforcing correlations such that the thermodynamic constraints are satisfied.\par
\subsection{Specific Example}
In this work we take the natural variables $\mathbf{x}=(lnT,lnV)^T$ and limit our EOS outputs to the pressure and internal energy, $\mathbf{y}=(P,U)^T$. The transformation operator comes directly from the thermodynamic relations \eqref{eq:ops},
%
%
which following log-transforming the input points become
\begin{equation}
  \mathcal{\hat{Y}}_{\mathbf{x}}\equiv\mathcal{\hat{Y}}_{(lnT,lnV)^T} =
  \begin{pmatrix}
    -\frac{1}{V}\frac{\partial}{\partial lnV} \\
    1 - \frac{\partial}{\partial lnT}
  \end{pmatrix}
  {\rm .}
\end{equation}
The log transformation of inputs allows for a more numerically stable handling of the wide range of thermodynamic conditions for whith the EOS model must be applied. The outputs also vary over a wide range, however log scaling of the outputs results in a non-linear operator $\mathcal{\hat{Y}}_\mathbf{x}$, precluding our simple use of GPs. Instead we use a GP to model the excess free energy, $\Delta F(\mathbf{x}) = F(\mathbf{x}) / F_R(\mathbf{x}) \sim GP(1,\,K_{\Delta F}(\mathbf{x}, \mathbf{x}^\prime))$ and the EOS model becomes
\begin{equation}
  \frac{\mathbf{y}(\mathbf{x})}{\mathbf{y}_R(\mathbf{x})}\sim GP(\mathcal{\hat{Y}}_\mathbf{x}\,1,\,\mathcal{\hat{Y}}_{\mathbf{x}^\prime}K_{\Delta F}(\mathbf{x}, \mathbf{x}^\prime)\mathcal{\hat{Y}}_\mathbf{x}^T)
  ~{\rm ,}
\end{equation}
where $\mathbf{y}_R(\mathbf{x})$ are the EOS outputs for the reference free energy model.
\par
Using equations \eqref{eq:eos_gp_app} and \eqref{eq:eos_kernel_elements} results in a two dimensional matrix of kernel functions giving the physics-driven correlations between EOS quantities,
\begin{gather}
  \begin{align}
  K_{\Delta\mathbf{y}}(\mathbf{x}, \mathbf{x}^\prime) &\equiv \mathcal{\hat{Y}}_{\mathbf{x}^\prime}K_{\Delta F}(\mathbf{x}, \mathbf{x}^\prime)\mathcal{\hat{Y}}_\mathbf{x}^T \\
  &=
  \begin{pmatrix}
    k_{PP}(\mathbf{x},\mathbf{x}^\prime) & k_{PU}(\mathbf{x},\mathbf{x}^\prime) \\
    k_{UP}(\mathbf{x},\mathbf{x}^\prime) & k_{UU}(\mathbf{x},\mathbf{x}^\prime) \\
  \end{pmatrix}
  {\rm ,}
\end{align}
  \label{eq:eos_kernel_structure}
\end{gather}
where, for example, the kernel for correlations between pressure values is given by
\begin{equation}
  \begin{aligned}
  k_{PP}(\mathbf{x},\mathbf{x}^\prime) &\equiv
  \mathcal{\hat{Y}}_{\mathbf{x}^\prime}^{(P)}\, \mathcal{\hat{Y}}_{\mathbf{x}}^{(P)}\,K_{\Delta F}(\mathbf{x}, \mathbf{x}^\prime) \\
  &= \frac{1}{V^\prime}\frac{\partial }{\partial lnV^\prime} \frac{1}{V}\frac{\partial}{\partial lnV} K_{\Delta F}(lnT, lnV; lnT^\prime, lnV^\prime)
  ~{\rm .} \label{eq:kernel_functions_matrix}
\end{aligned}
\notag
\end{equation}
\par
For our choice of inner kernel function $K_{\Delta F}(lnT, lnV; lnT^\prime, lnV^\prime)$ given in equation \eqref{eq:rbf_kernel}, we evaluate the 4 resulting EOS kernels using the symbolic algebra package \texttt{sympy}. This allows an easy derivation of all the required terms to implement our multi-output GP, as well as gradients to accelerate the optimization of hyperparameters.\par
\begin{figure*}
\centering
\includegraphics[scale=1]{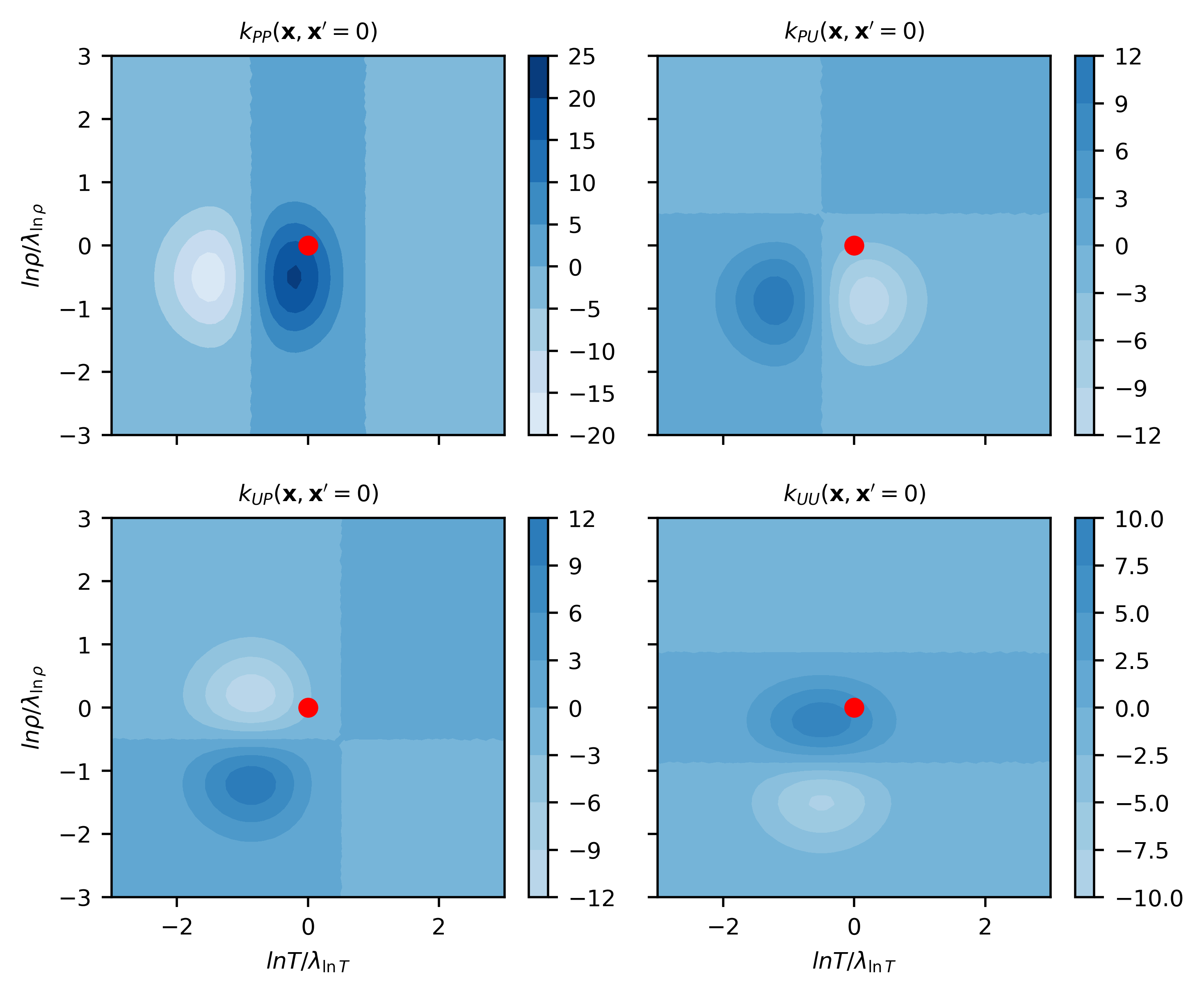}
\caption{The correlation structure in the outputs of our Gaussian Process, induced by the thermodynamic constraints. We plot the degree of correlation between outputs at the red point and a test point, as a function of test point location. Panels are labeled with the output quantity at the test point first, i.e., the "Pressure-Energy"  panel shows the correlation between the pressure at the test point and the energy at the red point. The asymmetric correlation structure, and correlations between different EOS quantities, are a direct result of the thermodynamic consistency of our approach.}
\label{fig:kernel}
\end{figure*}

We plot the resulting kernel functions in figure \ref{fig:kernel}. This plot shows each of the elements of equation \eqref{eq:kernel_functions_matrix} as a function of $\mathbf{x}$ and with $\mathbf{x}^\prime=(0,0)^T$ shown in red. The thermodynamic constraints result in asymmetric correlations between EOS quantities with important implications for the mutual information between different EOS quantities at the same and different points.

\end{document}